\documentclass[nofootinbib,a4paper,aps,pra,showpacs,preprintximbers,twocolumn]{revtex4-2}
\usepackage{cmap}
\usepackage{graphics,graphicx,epsfig}
\usepackage{epstopdf}
\usepackage[centertags]{amsmath}
\usepackage{amsfonts}
\usepackage{amssymb}
\usepackage[utf8]{inputenc}
\usepackage[english]{babel}
\usepackage{graphicx}
\usepackage{amsthm,color}
\usepackage{euscript}
\usepackage{indentfirst}
\usepackage{xcolor}
\usepackage{braket}
\usepackage{type1ec}%
\usepackage{physics}
\usepackage{tabularx}
\usepackage[T2A]{fontenc}	
\DeclareGraphicsExtensions{.pdf,.png,.jpg,.eps}

\begin{document}
	\def\BY{\begin{eqnarray}}
	\def\EY{\end{eqnarray}}
	\def\L{\label}
	\def\nn{\nonumber}
	\def\ds{\displaystyle}
	\def\({\left (}
	\def\){\right )}
	\def\[{\left [}
	\def\]{\right]}
	\def\<{\langle}
	\def\>{\rangle}
	\def\h{\hat}
	\def\td{\tilde}
	\def\r{\vec{r}}
	\def\ro{\vec{\rho}}
	\def\h{\hat}
	\preprint{APS/123-QED}
	
\title{\textbf{On the gauge-invariant orbital angular momentum of vector optical vortices in the paraxial approximation and beyond}}
    \author{E. V. Maliutina}
    \author{E. A. Vashukevich}
     \author{ V. A. Leonov}
      \author{D. D. Reshetnikov}
        \author{T. Yu. Golubeva}
    
\affiliation{%
	Saint-Petersburg State University, Universitetskaya Nab. 7/9, St. Petersburg 199034, Russia
}%
	\begin{abstract}

In this paper we present a gauge-invariant formalism for describing the spin and orbital angular momenta of scalar and vector optical vortices. The correction term, arising from consistent incorporation of the transverse vector potential's nonlocality, restores gauge invariance and makes it possible to separate the total angular momentum into spin and orbital components. We propose a method for calculating the orbital and spin angular momenta of multimode vector vortex fields and show that the orbital angular momentum of such beams contains two qualitatively different components. The polarization-independent contribution to the orbital angular momentum is related to the topological charge that determines the order of the phase singularity, while the polarization-dependent one is proportional to the helical topological charge that determines the spatial inhomogeneity of the polarization. We also discuss the quantization procedure for fields with spatially inhomogeneous polarization in terms of modes with a helical topological charge, and examine the limits of applicability of the proposed formalism.

\end{abstract}
\pacs{42.50.Dv, 42.50.Gy, 42.50.Ct, 32.80.Qk, 03.67.-a}
\maketitle

\section{Introduction}

Optical vortex fields can be divided into two classes, namely scalar and vector vortices, according to the nature of their phase and polarization spatial profiles. Scalar optical vortices can be defined as structured light fields that have a phase singularity in a transverse plane and thus carry orbital angular momentum (OAM). Such fields have a spiral wavefront characterized by the phase factor $e^{il\phi}$, where $l$ is the topological charge and $\phi$ is an azimuthal angle. Since the topological charge can take any integer value, optical vortices provide a degree of freedom for encoding information, which makes them promising for modern optical and quantum technologies  \cite{Mirhosseini2015, Bouchard2018}, in particular, for discrete-variable quantum computation.

The high dimensionality of the Hilbert space of OAM states makes it possible to implement more complex quantum algorithms and perform operations with higher information capacity per single photon \cite{Erhard2018}. Moreover, it has been demonstrated that OAM-entangled states can be produced with high fidelity \cite{Dada2011, VahukevichSq}, opening the way to the development of scalable quantum computing systems. Recently, several implementations of entangling and nonlocal logic gates for OAM qubits have been proposed in Refs. \cite{Vashukevich2024a, Bashmakova2024}. Optical vortices are also being actively explored as information carriers in quantum memory. In this context, OAM provides an additional degree of freedom, which makes it possible to store and retrieve multidimensional quantum states \cite{Ding2015, Vashukevich2020}.

It was first demonstrated in Ref. \cite{Allen1992} that beams with a phase singularity carry orbital angular momentum. For scalar paraxial beams with spatially uniform polarization structure, the projection of the total angular momentum onto the beam propagation axis was found to satisfy $J_z \propto l + \sigma$ where $l$ is the topological charge and $\sigma$ is the helicity, defined as the normalized difference between intensities of the left- and right-circularly polarized components. However, the question of whether total angular momentum of scalar vortices can be separated into independent observable spin and orbital parts remains open.

In Ref. \cite{vanEnk1994}, it was shown that using the gauge-invariant transverse vector potential $\boldsymbol{\mathcal{A}}$, one can formally define gauge-invariant operators $\hat{L}^{inv}$ and $\hat{S}^{inv}$ for scalar vortex fields. However, these operators do not satisfy the commutation relations typical for angular momentum operators:
\BY
    &&[\hat{S_k}, \hat{S_j}]= i \hbar \varepsilon_{kjn}\hat{S}_n,\\
    &&[\hat{L}_k, \hat{L}_j]= i \hbar \varepsilon_{kjn} \hat{L}_n.
\EY
Instead, the following commutation relations hold:
\BY
 &&[\hat{S}_k^{inv}, \hat{S}_j^{inv}]=0,\\
 &&[\hat{L}_k^{inv},
  \hat{L}_j^{inv}]= i \hbar \varepsilon_{kjn} (\hat{L}_n^{inv}+\hat{S}_n^{inv})
\EY

In the expressions above and throughout the paper we assume summation over all repeated indices. Moreover, since we deal with vectors in Euclidean space, we do not distinguish between co- and contravariant quantities for simplicity write all indices as subscripts.
Strictly speaking, the derived operators cannot be interpreted as the orbital and spin angular momentum operators of a photon.

Nevertheless, in the paraxial approximation this separation becomes physically meaningful. For paraxial beams (e.g., Laguerre–Gaussian modes), the spectra of the operators become discrete, and the eigenvalues are integer multiples of $\hbar$ \cite{vanEnk1994}. However, physical sense can be attributed only to the components of the orbital and spin angular momenta that are directed along the propagation axis, that is, $L_z$ and $S_z$. These components remain unchanged as light propagates in vacuum \cite{vanEnk1994} and can also be measured independently in experiments \cite{Leach2004}. Finally, they are transferred to matter upon interaction \cite{ONeil2002}.

Besides scalar optical vortices, there exists a broader class of structured light fields known as vector optical vortices, or beams with a spatially inhomogeneous polarization structure \cite{Rosales2018}. Unlike scalar vortices, in which the polarization is uniform across the beam cross-section and the topological charge determines only the phase structure, in vector vortices the direction of the polarization vector is not constant but varies in the transverse plane, creating polarization singularities \cite{Zhan2009}.

As shown in Ref. \cite{Forbes2021}, vector vortices are a natural extension of scalar vortices and provide new ways of light control. The inhomogeneous polarization structure and phase singularity of vector beams allow us to describe quantum states of such fields in a Hilbert space of higher dimension than that of scalar beams. This makes them particularly appealing for applications in quantum optics and information transfer \cite{Ndagano2018}.

Due to their spatially inhomogeneous polarization structure, vector vortex beams exhibit greater robustness against distortions in a turbulent atmospheric channel, as shown in Refs.\cite{Ndagano2018, Abolhassani2025, Nape2019}. Furthermore, polarization inhomogeneity enables the design of protocols invariant under rotations of the transmitter and receiver \cite{reshetnikov2025}.

For vector optical vortices no gauge-invariant separation of the total angular momentum into orbital and spin parts has yet been achieved — either classically or quantum-mechanically. In these fields, polarization and phase are closely interwoven: angular dependence enters both the phase and the direction of the polarization vector.

In the present work, we attempt to construct gauge-invariant projections of the angular momenta both in the paraxial approximation and beyond it. Starting from the requirement of gauge invariance of the vector potential, we show that gauge-invariant quantities can be constructed only beyond the paraxial case, in the form of asymptotically convergent series, because of the nonlocality of the vector potential.

We also calculate the gauge-invariant OAM for both scalar and vector fields. Our results show that two parameters of distinct physical nature contribute to the OAM of vector vortex fields: one accounts for the phase inhomogeneity, and the other for the spatially inhomogeneous polarization structure. In addition, the sign of the polarization-dependent contribution is determined by the quantum number associated with the spin angular momentum. We also discuss in detail the experimental measurement of the individual OAM contributions and the origin of the corrections required by gauge invariance.

\section{Gauge-invariant vector potential}

In this section, we describe the classical theory of angular momentum, which provides the basis for the construction of the gauge-invariant operators of orbital and spin angular momenta in quantum theory.

The total angular momentum of the electromagnetic field is defined as the volume integral:
\BY
   && J^{inv}_i = \int \(r_j\overline{E}_i B_j-r_j\overline{E}_jB_i\) \, d\bold{r},
    \label{eq:J_total}
\EY
where $r_i$ is a three-dimensional spacial coordinate, the bar over a vector component denotes complex conjugation, and bold symbols denote vectors. Here and below, the superscript $inv$ indicates the gauge-invariance of a quantity. The total angular momentum $\bold{J^{inv}}$ is a gauge-invariant quantity and does not change under a gauge transformation. Using the relation between the magnetic field and the vector potential and integrating by parts, the expression for $\bold{J^{inv}}$ can be transformed into the following form \cite{barnett2016}:
\BY
&&J_i = \int \left[ \varepsilon_{ijk}r_j \overline{E_n} \partial_k A_n \;+\;\varepsilon_{ijk} \overline{E_j} A_k  \right] d\bold{r},
\label{eq:J_split}
\EY
This definition, however,  is generally not gauge-invariant, since the expressions for the spin and orbital angular momentum densities depend explicitly on the vector potential $\mathbf{A}$, which is defined only up to a gradient of an arbitrary scalar function. The first term is interpreted as a component of the orbital angular momentum $L_i$, and the second as a component of the spin angular momentum $S_i$, with densities given as follows:
\BY
  && \mathcal{L}_i=\varepsilon_{ijk}r_j\overline{E_n} \partial_k A_n, \\
    &&\mathcal{S}_i=\varepsilon_{ijk} \overline{E}_j A_k.
\EY
Under a gauge transformation, the contributions to the total angular momentum can be redistributed between these two densities, while the fields $\bold{E}$ and $\bold{B}$ themselves remain unchanged. Therefore, separating the total angular momentum into orbital and spin parts requires a gauge-invariant approach. In other words, it is necessary to distinguish the gauge-independent part of the vector potential.

According to the Helmholtz theorem, the vector potential $\bold{A}$ can be represented as the sum of a transverse (divergence-free) and a longitudinal (curl-free) component:
\BY
&&A_i = \mathcal{A}_i +\partial_i \psi,
\label{eq:helmholtz}
\EY
where $\psi$ is a scalar potential. The transverse component of the potential is divergence-free:
\BY
&&\partial_i \mathcal{A}_i= 0.
\label{eq:div_rot_conds}
\EY

A gauge transformation of the electromagnetic field can be performed by adding the gradient of an arbitrary function to the vector potential and does not change the transverse component $\boldsymbol{\mathcal{A}}$ . This component can therefore be considered gauge-invariant and, thus, insensitive to the contribution of potential electrostatic fields. Performing a gauge transformation from $\mathbf{A}$ to $\boldsymbol{\mathcal{A}}$, we can uniquely express the electric field $\mathbf{E}$ in terms of $\boldsymbol{\mathcal{A}}$ and as a result, obtain expressions for the orbital and spin angular momenta \cite{barnett2016}. These expressions depend only on the gauge-invariant part of the vector potential:
\BY
&&\mathcal{L}_i^{inv}\equiv\mathcal{L}_i\[\mathcal{A},\mathcal{A}\]=-\varepsilon_{ijk}r_j\overline{(\partial_0\mathcal{A}_n)} (\partial_k\mathcal{A}_n),
\label{eq:L_invariant}\\
&&\mathcal{S}^{inv}_i\equiv\mathcal{S}_i\[\mathcal{A},\mathcal{A}\] =-\varepsilon_{ijk}\mathcal{A}_k \overline{(\partial_0\mathcal{A}_j)} .
\label{eq:S_invariant}
\EY
Here, $\partial_0$ denotes the time derivative. For convenience in subsequent calculations, we introduce the notation $\mathcal{L}_i[\mathcal{A},\mathcal{A}]$ and $\mathcal{S}_i[\mathcal{A},\mathcal{A}]$. The first argument denotes the function differentiated with respect to time. The second argument denotes either the function differentiated with respect to the spatial coordinates, as in Eq. (11), or a function that is not differentiated at all (as in Eq. (12)). We now determine the gauge-invariant part of the vector potential $\mathbf{A}$. In order to do this, we use definition \eqref{eq:div_rot_conds} and calculate the divergence of the full potential. This yields an equation for the function $\psi$:
\BY
&&\partial_i A_i= \partial_j(\partial_j\psi).
\label{eq:poisson_f}
\EY
Thus, the problem of finding the gauge-invariant transverse potential reduces to solving the Poisson equation (\ref{eq:poisson_f}), which gives the formal expression for the gradient of $\psi$:
\BY
  && \partial_s\psi= \partial_s \left(\{\partial_j\partial_j\}^{-1}(\partial_i A_i) \right),
    \label{eq:Aparl_formal}
\EY
Here, $\{\partial_j\partial_j\}^{-1}$ denotes the inverse Laplace operator. To calculate this inverse operator explicitly, it is convenient to work in Fourier space. The Fourier transform of the longitudinal (curl-free) component of the vector potential has the form:
\BY
   &&\mathcal{F}[\partial_s\psi] = k_s \left( \frac{ k_j\mathcal{F}[A_j]}{k^2} \right),
    \label{eq:Aparl_fourier}
\EY
where $\mathcal{F}[A_j]=\int A_j e^{-i k_s r_s}\,d\mathbf{r}$ is the Fourier transform of the full potential $\mathbf{A}$. The quantities $k_j$ are the components of the wave vector, and $k$ is the wave number.

Let us assume that the vector potential describes a wave propagating along the $z$-axis with wave number $k_0=\frac{\omega_0}{c}$, where $\omega_0$ is the carrier frequency:
\BY
  && \bold{ A} (\bold{r}_\perp,z)=\bold{ A}_0 (\bold{r}_\perp) e^{i k_0 z}.
\EY
Here, $\mathbf{r}_\perp$ is the radius vector in the plane perpendicular to the propagation axis. We neglect the dependence of $\mathbf{A}_0$ on the coordinate $z$ because in the slowly varying amplitude approximation, we may always assume that $\partial_z\mathbf{A}_0(\mathbf{r}_\perp,z)\ll i k_0$. The Fourier transform of such a field is given by:
\BY
   && \bold{A}(\bold{k}) = \bold{A_0}(\bold{k}_{\perp})\delta(k_0 - k_z).
    \label{furier}
\EY
It contains a delta function that fixes the component $k_z$ of the wave vector $\mathbf{k}$. In Eq. \eqref{furier}, $\mathbf{A}_0(\mathbf{k}_{\perp})$ is the Fourier transform of $\mathbf{A}_0(\mathbf{r}_\perp)$.

We can expand the quantity $\frac{1}{k^2}$ in \eqref{eq:Aparl_fourier} in a series in the small parameter $\frac{k_\perp^2}{k_z^2}$, taking into account that in the paraxial approximation we may assume $k_\perp \ll k_z$. Since $k_z$ is fixed and equal to $k_0$, we obtain:
\BY
  &&  \frac{1}{k^2} = \frac{1}{k_0^2 + k_\perp^2} = \frac{1}{k_0^2} \left(1 - \frac{k_\perp^2}{k_0^2} + \frac{k_\perp^4}{k_0^4} - \cdots \right).
    \label{eq:k2_expansion}
\EY
By substituting \eqref{eq:k2_expansion} into \eqref{eq:Aparl_fourier}, one can perform the inverse Fourier transform and obtain:
\BY
  && \{\partial_i\partial_i\}^{-1} = -\frac{1}{k_0^2} \sum_{n=0}^{\infty} \left( \frac{\{\partial_j\partial_j\}^\perp}{k_0^2} \right)^n,
    \label{eq:laplacian_inv_exp}
\EY
where $\{\partial_j\partial_j\}^\perp$ is the transverse part of the Laplace operator.

Inserting this expansion into expression \eqref{eq:Aparl_formal} and taking into account Eq. \eqref{eq:helmholtz}, we obtain a general formula for the transverse part of the vector potential in the form of a series:
\BY
&&    \mathcal{A}_s = A_s+\frac{1}{k_0^2}\partial_s\sum_{n=0}^{\infty} \left( \frac{\{\partial_j\partial_j\}^\perp}{k_0^{2}} \right)^n  \partial_i A_i.
    \label{eq:Aparl_series_general}
\EY
Expression \eqref{eq:Aparl_series_general} is valid for arbitrary electromagnetic waves whose amplitude varies slowly along the propagation axis. It makes it possible to calculate the gauge-invariant vector potential beyond the paraxial approximation to any prescribed accuracy.

In a number of studies, for example, in Refs. \cite{bliokh2015, berry2009}, a correction to only one component of the vector potential is introduced to achieve gauge invariance:
\BY
   &&\mathcal{A}_z = \frac{i}{k_0} \partial_j A_j.
    \label{eq:Az_perp}
\EY
Eq. \eqref{eq:Az_perp} agrees with Eq. \eqref{eq:Aparl_series_general} to the zeroth order in $\frac{\{\partial_j\partial_j\}^\perp}{k_0^2}$, that is, when only the term with $n=0$ is taken into account.However, we retain all terms of the series in our calculations. As shown below, the requirement of gauge invariance requires the vector potential to be strongly nonlocal. This must be taken into account when calculating physical quantities.

Once the divergence-free component of the potential has been constructed and Eq. \eqref{eq:Aparl_series_general} has been substituted into Eq. \eqref{eq:L_invariant}, one can write an expression for the gauge-invariant orbital angular momentum density. In the resulting expression, several terms of distinct physical nature can be identified. These terms depend, respectively, on the original potential $\mathbf{A}$, on a combination of the longitudinal component and $\mathbf{A}$, and solely on the longitudinal component:
\BY
&&\mathcal{L}^{inv}_i=\mathcal{L}_i\[A,A\]+\mathcal{L}_i\[A,\partial_j\psi\]+\mathcal{L}_i\[\partial_s\psi,\partial_j\psi\].
\EY

A similar expression can be written for the spin angular momentum density. It is important to note that if we consider the expression for the total angular momentum density $\mathcal{J}$, defined as the integrand in Eq. \eqref{eq:J_total}, and rewrite it in terms of the transverse component of the potential, then the longitudinal component, defined as the gradient of a scalar potential, does not contribute to the magnetic field. Thus, the expression for $\mathcal{J}$ will contain no terms that depend only on the longitudinal component:
\BY
\mathcal{J}^{inv}_i\equiv\mathcal{J}_i\[\mathcal{A},\mathcal{A}\]=\mathcal{J}_i\[A,A\]+\mathcal{J}_i\[A,\partial_j \psi\]\label{J_INV}.
\EY

Thus, when the gauge-invariant potential is used, the equality $\mathcal{J}^{inv}_i = \mathcal{L}^{inv}_i + \mathcal{S}^{inv}_i$ no longer holds exactly. This makes the transition from Eq. \eqref{eq:J_total} to Eq. \eqref{eq:J_split} incorrect, unless the contributions $\mathcal{L}_i[\partial_s\psi,\partial_j\psi]$ and $\mathcal{S}_i[\partial_s\psi,\partial_j\psi]$ are identically zero. As we show below, this inconsistency arises from the nonlocal definition of the gauge-invariant potential. A naive calculation then leads to physically incorrect contributions to the orbital and spin angular momenta.

The physically incorrect terms can be removed if the transformation from Eq. \eqref{eq:J_total} to  Eq. \eqref{eq:J_split} is carried out more rigorously. The key point is to include a surface integral that is usually taken as zero. However, the nonparaxial corrections to $\mathcal{A}$ produce a nonzero surface term when integrating over an arbitrary finite volume:
\BY
 && J^{inv}_i = \int \(r_j\overline{E}_i B_j-r_j\overline{E}_jB_i\) \, d\bold{r}=\nn\\ 
 &&L_i+S_i-\oint_{\partial V} \varepsilon_{ijk}\, r_j\, \overline{E}_l\, A_k\, n_l \, d^2 r,\label{surf}
\EY
where $n_l$ denotes the component of the outward normal. The series in Eq. \eqref{eq:Aparl_series_general}, which describes vortex potentials, can be summed analytically, but it converges only asymptotically, as shown in Section III. Consequently, the surface integral cannot be neglected at finite scales. If we convert the surface integral in Eq. \eqref{surf} into a volume integral, we can follow the same reasoning as in Eq. \eqref{eq:L_invariant} and define a gauge-invariant density for the correction term:
\BY
\mathcal{G}^{inv}_i\equiv\mathcal{G}_i\[\mathcal{A},\mathcal{A}\]\equiv\varepsilon_{ijk}\overline{(\partial_0\mathcal{A}_n)}(\partial_n(r_j\mathcal{A}_k)).\label{G_INV}
\EY
In this case, to any scale, that is, for any finite number of terms of the series taken into account, one can write an exact equality

\BY
\mathcal{J}^{inv}_i=\mathcal{L}^{inv}_i+\mathcal{S}^{inv}_i-\mathcal{G}^{inv}_i.\label{conserv}
\EY
Note that this correction term is essentially the Belinfante correction \cite{Benlinfante40,Bliokh2014}, which restores gauge invariance of the total angular momentum. In the next section, we show that representing the vector potential as a superposition of scalar or vector vortex fields yields an analytical expression for the transverse potential. We also demonstrate that the correction introduced above eliminates the unphysical contributions to the angular momentum densities. This makes it possible to calculate gauge-invariant physical quantities at any scale.

\section{Spin and orbital angular momenta for the gauge-invariant potential of vortex fields}

In this section, we apply the formalism proposed in the previous section to construct the gauge-invariant vector potential. We use it to calculate the spin and orbital components of the total angular momentum for two types of electromagnetic fields, scalar vortices, which are polarization-homogeneous, and vector vortices, which are polarization-inhomogeneous.

\subsection{Scalar vortices}
The vector potential of a scalar vortex field can be represented as the product of two independent factors. The first factor is the polarization vector, written as a superposition of right and left circular polarizations with certain coefficients. The second factor is the amplitude-phase function, which carries information about the order of the phase singularity typical of vortex fields. Without loss of generality, we assume that the amplitude-phase part can be represented as a Laguerre-Gaussian function. The order of the phase singularity of the field is given by the topological charge $l$:
\BY
&&\bold{A}^{sc}(\rho, \phi)= \left(\alpha_R \bold{e}_R
 +\alpha_L \bold{e}_L
\right) U_l(\rho) e^{i l \phi},
\label{eq:pot_vortex2}
\EY
where $\mathbf{e}_{R,L} = \frac{1}{\sqrt{2}}(1, \pm i, 0)^\mathsf{T}$ are the circular polarization vectors. The function $U_l(\rho)=\sqrt{\frac{2}{{\pi \rho_0^2 l! }}}e^{-\rho^2/\rho_0^2}(\frac{\sqrt2 \rho}{\rho_0})^{l}$ is the amplitude part of the Laguerre-Gaussian mode $\mathrm{LG}_0^l$, which does not depend on the longitudinal coordinate $z$. This approximation appears justified, since we treat the amplitude as slowly varying and consider the beam near its waist, with waist width $\rho_0$. The complex coefficients $\alpha_R$ and $\alpha_L$ determine the amplitude and phase of each polarization component. In Eq. (\ref{eq:pot_vortex2}), the absence of the index $l$ in the coefficients $\alpha_R$ and $\alpha_L$ indicates that the vortex field is scalar. In other words, the polarization structure and the phase structure in the expression above factorize.

Here and below, we consider fields in free space. For brevity, we omit the rapidly oscillating spatiotemporal factor $e^{i k_0 z - i \omega_0 t}$. That is, we assume that
 \BY
 &&\bold{A}^{sc}(\rho, \phi,z)=\bold{A}^{sc}(\rho, \phi)e^{i k_0 z - i \omega_0 t}.
 \EY

This representation of the vortex potential is particularly convenient. The analytical structure of the functions $U_l(\rho)$ allows the terms of the series in Eq. \eqref{eq:Aparl_series_general} to be calculated explicitly:
\BY
  &&  \left( \frac{\{\partial_j\partial_j\}^\perp}{k_0^{2}} \right)^n  \partial_i A^{sc}_i(\rho,\phi) = \nn\\
      && \frac{(-4)^n \sqrt{2}}{k_0^{2n}}\Biggl(-\frac{n! \,  \, e^{2i\phi} \, \rho \, L_n^{(l+1)}\!\left(\frac{\rho^2}{\rho_0^2}\right) \alpha_R}{\rho_0^{2(n+1)}} \nn+ \\
      &&  \frac{(n+1)! \, \,  \, L_{n+1}^{(l-1)}\!\left(\frac{\rho^2}{\rho_0^2}\right) \alpha_L}{\rho \, \rho_0^{2n}}\Biggr)U_l(\rho) e^{i l \phi},
      \label{n_laplacian}
\EY
where $L_l^\alpha(x)$ are the generalized Laguerre polynomials. The $n$-th order correction term to the vector potential is obtained from Eq. \eqref{n_laplacian} by taking the gradient, in accordance with Eq. \eqref{eq:Aparl_series_general}. The radial derivatives of the Laguerre polynomials arising here are calculated using the well-known recurrence relation
\BY
&&dL_n^{(\alpha)}(x)/dx = -L_{n-1}^{(\alpha+1)}(x).
\EY
Thus, using Eq. \eqref{eq:Aparl_series_general}, one can obtain an analytical expression for the gauge-invariant vector potential of a scalar vortex field $\boldsymbol{\mathcal{A}}^{sc}$. Next, for calculating the spin and orbital angular momenta of scalar fields, we use Eqs. \eqref{eq:L_invariant} and \eqref{eq:S_invariant}. Here, unlike previous studies, we do not truncate the series obtained from the expansion in Eq. \eqref{eq:laplacian_inv_exp}, and, in this way, go beyond the paraxial approximation. Since we have an analytical expression for the potential, and therefore for the angular momentum densities, one can integrate the $z$-components of the gauge-invariant angular momentum densities over all space explicitly. As a result, one can obtain the projections of the spin and orbital angular momenta onto the propagation axis as asymptotic series in the nonparaxiality parameter $\xi$:
\BY
 &&   S^{inv}_z \propto \sigma\left(1 + \sum_{n=0}^{\infty} \frac{(-1)^{n+1} (1 + |l| + n)!\xi^{n+1} }{  |l|!}\right), \label{Spin}\\
&& L^{inv}_z \propto l\left(1 +  \sum_{n=0}^{\infty}  \frac{(-1)^{n+1} (1 + |l| + n)!\xi^{n+1} }{|l|!}\right)\nn\\
&&+\frac{l+\sigma}{2}\sum_{n=0}^{\infty}\frac{(1+|l|+2n)! (1+(2+|l|+2n)\xi)\xi^{2n+1}}{|l|!} ,\;\;\;\;\;\;\;
\label{OAM}
\EY
where $\sigma = |\alpha_R|^2 - |\alpha_L|^2$ is the helicity ($\sigma = \pm 1$ for right and left circular polarization, respectively), and $\xi=\frac{\sqrt{2}}{k_0^2\rho_0^2}$ is the nonparaxiality parameter. The chosen parameter is small in the expansion. However, the factorial growth of the coefficients multiplying $\xi$ makes the series divergent in the usual sense. Strictly speaking, it converges only asymptotically. In the expressions above, the first term appears naturally from the integration of $\mathcal{S}_z[A^{sc},A^{sc}]$ and $\mathcal{L}_z[A^{sc},A^{sc}]$. The alternating series in the nonparaxiality parameter arises from the integration of the density components $\mathcal{S}_z[A^{sc},\partial_s\psi^{sc}]$ and $\mathcal{L}_z[A^{sc},\partial_s\psi^{sc}]$. These contributions turn out to be the same for both the spin and orbital angular momenta, up to the replacement of $l$ by $\sigma$. It is noteworthy that when the terms depending only on the gradient of the scalar potential are integrated, the contribution to the spin part is found to be zero. In contrast, $\mathcal{L}_z[\partial_j\psi^{sc},\partial_s\psi^{sc}]$ produces a sign-definite term, which also contains the spin-orbit factor $l+\sigma$.

To give physical meaning to the terms represented by asymptotic series, we use the Borel regularization method \cite{BorelReg}. This method is well suited for summing alternating series. It allows one to obtain a function of $\xi$ that approximates the sum of the series when an optimal number of terms is considered. That is, for any sufficiently small finite value of $\xi$, one can see that the partial sums approach the resulting function up to an optimal number $N$, and then begin to diverge. Conversely, for any finite number of summed terms, one can choose a value of $\xi$ such that the truncated series approximates the function obtained. The sign-definite term in Eq. \eqref{OAM} can also be regularized. However, it has an unphysical pole in the Borel image and diverges at certain values of the nonparaxiality parameter. Thus, a direct naive calculation of the orbital angular momentum projection using the gauge-invariant part of the vector potential gives a counterintuitive result, since the appearance of a Borel pole in this case has no clear physical interpretation.
	
We now show that the calculated correction series can be augmented, and the unphysical contributions to the angular momentum can be eliminated. Expressions for the total angular momentum and for the correction term can be obtained by integrating the densities $\mathcal{J}^{inv}_z$ and $\mathcal{G}^{inv}_z$, given by Eq. \eqref{J_INV} and Eq. \eqref{G_INV}, respectively:
\BY
&&   J^{inv}_z \propto \frac{l+ \sigma}{2}\left(1+\sum_{n=0}^{\infty}(-1)^{n} \frac{(|l| + n)!\xi^{n}}{|l|!}\right)\label{Glob}\\
&&+\frac{\sigma}{2}\sum_{n=0}^{\infty} \frac{(-1)^{n+1} (|l| + n)!((1+n)(1+|l|+n)\xi+n)\xi^{n} }{  |l|!},\nn\\
&& G^{inv}_z \propto  \frac{l+ \sigma}{2}\Bigg(1+ \nn\\
&&\left.\sum_{n=0}^{\infty} \frac{(-1)^{n+1} (|l| + n)!(1+2(1+n +|l|)\xi)\xi^{n} }{  |l|!}\right) \nn\\
&&+ \frac{ \sigma}{2}\sum_{n=0}^{\infty} \frac{(-1)^{n} (|l| + n)!((1+n)(1+|l|+n)\xi+n)\xi^{n} }{  |l|!}\nn\\
&&+  \frac{l+ \sigma}{2}\sum_{n=0}^{\infty}\frac{(1+|l|+2n)! (1+(2+2n+|l|)\xi)\xi^{2n+1}}{ |l|!}.\;\;\;\;\;\;\;
\label{Belin}
\EY

It is readily seen that the expression for the total angular momentum contains no sign-definite series generated by the angular momentum density associated with the scalar potential. The alternating series arising from $\mathcal{J}_z[A^{sc},\partial_j\psi]$ does not coincide with those for $L^{inv}_z$ and $S^{inv}_z$. This reflects the need to take the correction $G^{inv}_z$ into account when separating the total angular momentum into orbital and spin parts. The leading contribution (which contains no series) to the total angular momentum differs by a factor of two from the sum of the spin and orbital angular momenta. The sign-definite part in Eq. \eqref{Belin} is exactly the same as the sign-definite term in $L_z^{inv}$. Thus, with this correction and using Eq. \eqref{conserv} for the integrated quantities, we can eliminate the nonphysical contributions to the orbital angular momentum. Moreover, we can correct the term generated by the density $\mathcal{J}_z[A^{sc},\partial_j\psi]$, which turns out to be equal to $\mathcal{L}_z[A^{sc},\partial_j\psi]+\mathcal{S}_z[A^{sc},\partial_j\psi]-\mathcal{G}_z[A^{sc},\partial_j\psi]$.

We conventionally associate the contributions depending only on the topological charge $l$ with the orbital angular momentum, and those containing $\sigma$ with the spin angular momentum. Accordingly, the gauge-invariant angular momentum projections can be redefined as follows:
\BY
&&\tilde{L}^{inv}_z=\frac{l}{2}\left(1+\sum_{n=0}^{\infty} \frac{(-1)^{n} (|l| + n)!\xi^{n} }{  |l|!}\right) ,\label{L_Corr}\\
&&\tilde{S}^{inv}_z=\frac{\sigma}{2}\left(1+\sum_{n=0}^{\infty} \frac{(-1)^{n} (|l| + n)!\xi^{n} }{  |l|!} +\right.\nn\\
&&\left.\sum_{n=0}^{\infty} \frac{(-1)^{n} (|l| + n)!((1+n)(1+|l|+n)\xi+n)\xi^{n} }{  |l|!}\right).\label{S_Corr} \;\;\;\;\;\;\;
\EY
Now, for the redefined angular momenta, we can write an exact equality
\BY
&&J_z^{inv}=\tilde{L}^{inv}_z+\tilde{S}^{inv}_z.
\EY
It is of interest to trace the asymptotic behavior of the series in Eq. \eqref{L_Corr} and Eq. \eqref{S_Corr}. With the Borel regularization, the sums of the series can then be written in analytical form:
\BY
\sum_{n=0}^{\infty} \frac{(-1)^{n} (|l| + n)!\xi^{n} }{  |l|!}\equiv \frac{e^{1/\xi}}{\xi}\int_1^{\infty} e^{-\frac{t}{\xi}} \frac{dt}{t^{|l|+1}}.\label{Bsum}
\EY
The generalized exponential integral in Eq. \eqref{Bsum} converges absolutely for all positive real values of $\xi$. Thus, it makes it possible to calculate the correction to the angular momentum for any degree of nonparaxiality. The third term in the expression for spin angular momentum is identically zero for any $\xi$ in the sense of the Borel sum:
\BY
\sum_{n=0}^{\infty} \frac{(-1)^{n} (|l| + n)!((1+n)(1+|l|+n)\xi+n)\xi^{n} }{  |l|!}\equiv 0.\;\;\;\;
\EY
Figure 1 shows the orbital angular momentum of a scalar vortex normalized to the topological charge $l$, as a function of the inverse nonparaxiality parameter $1/\xi$, for several values of the topological charge $l$. 
\begin{figure}
	\includegraphics[scale=0.72
]{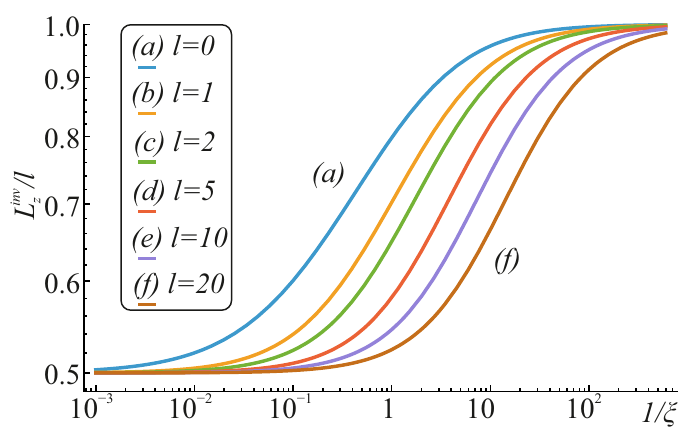}\caption{Dependence of the gauge-invariant orbital angular momentum on the inverse nonparaxiality parameter for various values of the topological charge $l$}
\end{figure}
For large values of the topological charge, smaller values of $\xi$ are required for good convergence to the paraxial limit. In this limit, the normalized angular momentum equals unity. This behavior is explained by the fact that the characteristic transverse size of the mode increases with the topological charge. Even small corrections taken into account therefore drive the system out of the paraxial regime. In the limit of an infinitely wide waist  (which we also call the paraxial limit) that is, as $\xi\rightarrow0$, the nonzero correction to the angular momenta tends to unity:
\BY
\lim\limits_{\xi\rightarrow0}\[\frac{e^{1/\xi}}{\xi}\int_1^{\infty} e^{-\frac{t}{\xi}} \frac{dt}{t^{|l|+1}}\]=1,
\EY
which ensures correct values of $L_z^{inv}$ and $S_z^{inv}$ in the paraxial limit:
\BY
&&\lim\limits_{\xi\rightarrow0}\[	L_z^{inv}\]=l\label{L_parax},\\
&&\lim\limits_{\xi\rightarrow0}\[	S_z^{inv}\]=\sigma\label{S_parax}.
\EY  

Therefore, this approach to constructing the gauge-invariant part of the vector potential and to calculating the physical quantities associated with it allows one to obtain the well-known result of Ref. \cite{Allen1992} without discarding any of the nonparaxial corrections, using only regularization of the series. It is also important that the Belinfante correction, although it does not allow one to separate the orbital and spin angular momentum densities explicitly, makes such a separation possible for integrated quantities. It thereby ensures correct values of the angular momenta and their gauge invariance. This result confirms the arguments given in Ref. \cite{Bliokh2014} about the impossibility of constructing local densities of the orbital and spin angular momenta separately. In the next section, we analyze the separation of the angular momentum of vector vortex fields and generalize the result to fields that are a superposition of vector vortices.

\subsection{Vector vortices}

Previously, we have considered scalar optical vortices, that is, fields with a spatially homogeneous polarization structure. Such fields can be described by a vector potential of the form given by Eq. \eqref{eq:pot_vortex2}. In these beams, the polarization is constant over the entire cross-section, and the topological charge $l$ fully determines the phase structure of the field. As shown above, for this class of fields the gauge-invariant formalism allows one to separate the orbital and spin components of the angular momentum. In the paraxial limit, these components satisfy $L_z \propto l$ and $S_z \propto \sigma$, where $\sigma$ is the helicity characterizing the handedness of the polarization.

However, there exists a broader class of vortex fields, namely vector optical vortices (i. e., beams with spatially inhomogeneous polarization). In these fields, the direction of the polarization vector is not constant but varies in the transverse plane, so the polarization and spatial structures are closely intertwined. The angular dependence is present not only in the phase but also in the direction of the polarization itself. In this section, we generalize the gauge-invariant formalism proposed above to the case of vector vortices.

For a vector vortex beam, there is a natural representation in the basis of radial and azimuthal polarizations, $\{\bold{e}_{RP}^{\;w}(\phi), \bold{e}_{AP}^{\;w}(\phi)\}$. According to the Jones formalism, the basis vectors are defined as:
\BY
\bold{e}^{\;w}_{RP}(\phi)= 
\begin{pmatrix}
\cos(w\phi) \\ \sin(w\phi) \\ 0
\end{pmatrix}, \quad 
\bold{e}^{\;w}_{AP}(\phi)=  
\begin{pmatrix}
-\sin(w\phi) \\ \cos(w\phi) \\ 0
\end{pmatrix}.\;
\label{basis}
\EY
The parameter $w$ in this representation characterizes the number of complete rotations made by the linear polarization vector upon going around the optical axis, as well as the direction of polarization rotation across the transverse plane. The parameter $w$ can take any integer value.

In this basis, an arbitrary field can be expressed as a superposition of radial (with subscript $RP$) and azimuthal (with subscript $AP$) components. We first consider the case of mode with a certain value of $w$, which is a superposition of two modes with the azimuthally symmetric polarization distribution. The vector potential then takes the following form:
\BY
&&\bold{A}^{vec}(\rho, \phi)
= U_w(\rho) \bigl( \alpha^w_{RP} \bold{e}^{\;w}_{RP}(\phi) + \alpha^w_{AP} \bold{e}^{\;w}_{AP}(\phi) \bigr).\;\;\;\;\;
\label{eq: Pot2}
\EY
For the mode represented by Eq. \eqref{eq: Pot2}, the intertwining of the polarization structure and spatial structure is manifested as follows. The angular dependence resides entirely in the basis vectors of radial and azimuthal polarization, $\mathbf{e}^{\;w}_{RP}$ and $\mathbf{e}^{\;w}_{AP}$, while the amplitude function $U_w(\rho)$, which describes the radial distribution of the field, is independent of the azimuthal angle $\phi$. As a result, differentiation with respect to $\phi$ affects only the polarization structure of the beam, and not its spatial amplitude. This means that a direct application of the operator $\hat{L}_z$ to the field in this representation does not permit an unambiguous separation of the contribution associated with the spatial phase distribution from those of the polarization structure.

To account for nonparaxial effects, we apply the gauge-invariant formalism proposed in the previous section. The expansion of the transverse vector potential $\mathcal{A}$ in the series in Eq. \eqref{eq:Aparl_series_general}, together with the approach described in detail in the previous section (see Appendix A), allows limiting values of the spin and orbital angular momenta to be written similar to those in Eq. \eqref{L_parax} and Eq. \eqref{S_parax}:
\BY
&&\lim\limits_{\xi\rightarrow0}\[	L_z^{inv}\]= w \sigma\label{Lvort},\\
&&\lim\limits_{\xi\rightarrow0}\[	S_z^{inv}\]=\sigma\label{Svort}.\EY  

It is worth noting that the nonparaxial corrections and their Borel images for the vector vortex are identical to those for a scalar vortex. The expressions above reflect the relationship between the spatial structure of the field and its polarization. The orbital angular momentum of such beams is proportional to the product of the helicity $\sigma$ and the helical topological charge $w$. This relationship is the reason for choosing the term "helical" to denote the topological charge $w$: it determines the contribution to the orbital angular momentum, which affected by the sign of the helicity. We also note that, for a mode with a specific value of $w$, the orbital angular momentum arises exclusively from the rotation of the basis. In contrast, differentiation of the amplitude with respect to the angle $\phi$ gives zero, thus, the orbital angular momentum of such fields originates only from the metric connection.

We now show that, for the simplest case of a multimode vector field, that is, a field containing two orthogonal polarization components with different topological charges $l_1$ and $l_2$, the expression for the gauge-invariant angular momentum contains two physically distinct contributions.
	
The vector potential of such a field, $\bold{A}^{tm}(\rho, \phi)$, can be represented by the expansion:
\BY
	\bold{A}^{tm}(\rho, \phi)=\bold{e}_R\alpha_R^{\;l_1}  e^{i l_1 \phi} U_{l_1}(\rho)
	+\bold{e}_L\alpha_L^{l_2} e^{i l_2 \phi} U_{l_2}(\rho).\;\;\;
	\label{eq:pot_vortex}
	\EY
We factor out the common phase factor in Eq. \eqref{eq:pot_vortex} to facilitate further calculations:
\BY
		\bold{A}^{tm}(\rho, \phi)&&=e^{i\frac{l_1+l_2}{2}\phi} \left(\bold{e}_R\alpha_R^{\;l_1}  e^{i \frac{l_1-l_2}{2} \phi}U_{l_1}(\rho) \right.\\
	&&\left.	+\bold{e}_L\alpha_L^{l_2}  e^{-i \frac{l_1-l_2}{2} \phi}U_{l_2}(\rho) 
		\right),
		\label{eq:pot_vortex1}
\EY
The different angular momenta of the left and right circular polarization components form a vector vortex whose polarization direction rotates around the propagation axis. In the particular case $l_1=l_2$, the expression reduces to a scalar vortex with a homogeneous polarization structure. The case $l_1=-l_2$ describes vector vortices with axially symmetric polarization and a helical topological charge $w=l_1$. In the general case, redefining $\frac{l_1+l_2}{2}\equiv l$ and $\frac{l_1-l_2}{2} \equiv w$, we can write the vector potential of such a field in the form:
\BY
  &&  \bold{A}^{tm}(\rho, \phi) =\nn\\&&= \frac{1}{\sqrt{2}}e^{i l \phi}\Bigl((\alpha_R^{l+w} U_{l+w}(\rho)+\alpha_L^{l-w}U_{l-w}(\rho))\bold{e}_{RP}^{\; w}(\phi)\nn\\&&+i(-\alpha_R^{l+w} U_{l+w}(\rho)+\alpha_L^{l-w}U_{l-w}(\rho))\bold{e}_{AP}^{\; w}(\phi)\Bigr).
\EY
This form clearly reveals the structure of the field. The factor $e^{i l \phi}$ accounts for the common, polarization-independent phase. The basis vectors $\mathbf{e}_{RP}^{\;w}$ and $\mathbf{e}_{AP}^{\;w}$ contain all the information about the polarization structure and are characterized by the helical topological charge $w$. The projections of the gauge-invariant angular momenta onto the propagation axis in the limit $\xi\rightarrow0$, calculated in the same way as in the previous cases, can then be written as follows
\BY
   && L^{inv}_z= l+w\sigma\label{OAMVec},\\
    &&S^{inv}_z= \sigma.
\EY
It can be noted that the orbital angular momentum of a vector vortex splits into two qualitatively different parts: a \textit{spin-dependent} part and a \textit{spin-independent} part. Importantly, the topological charges $l_1$ and $l_2$ do not contribute to $L_z^{inv}$ directly, and the orbital angular momentum is not simply the average of the angular momenta of the two modes. The first term is originated from the scalar phase singularity and associated with the result of differentiating the vector potential's amplitude itself. In contrast, the second term arises from differentiating the rotating basis vectors. Thus, the two topological charges $l$ and $w$ are naturally related to two physically distinct contributions to the orbital angular momentum.

Finally, we consider the most general case, namely an arbitrary multimode field with a vector potential $\mathbf{A}(\rho,\phi)$. We can expand the vector potential in terms of scalar vortices separately for each of the two orthogonal polarization components:
\BY
\bold{A}(\rho,\phi) = \sum_{s=-\infty}^{+\infty} \alpha_R^s \bold{e}_R   U_s(\rho) e^{is\phi}
	+ \sum_{l=-\infty}^{+\infty} \alpha_L^l \bold{e}_L   U_l(\rho) e^{il\phi}.\;\;\;\label{MultiA}
\EY
For such a field, it is convenient to consider the average topological charges in the right and left polarizations:
\BY
&&\langle l_{k}\rangle=\frac{\sum\limits_i l_i |\alpha_{k}^i|^2}{\sum\limits_j  |\alpha_{k}^j|^2}, k\in\{R,L\}.
\EY
In the case under consideration, the orbital angular momentum also contains polarization-dependent and polarization-independent contributions, as follows from Eq. \eqref{OAMVec}. Here, however, the helicity $\sigma$ and topological charges $l$ and $w$ are determined in terms of the average topological charges and the superposition coefficients as follows:

\BY
&&\sigma=\sum\limits_i(|\alpha_R^i|^2-|\alpha_L^i|^2),\\
&&l=\frac{1}{2}\(\langle l_R\rangle+\langle l_L\rangle\).\\
&&w=\frac{1}{2}\(\langle l_R\rangle-\langle l_L\rangle\).
\EY
That is, the helicity of the field is defined as the sum of the helicities in each mode with charge $l$, while the spin-dependent part is proportional to the difference of the average angular momenta in the orthogonal polarizations. It is noteworthy that that integer values of the topological charges $l$ and $w$ for a multimode field correspond to fields with axial symmetry of the phase and polarization, respectively. In contrast, fractional values appear solely as artifacts of averaging and are not related to the spectrum of the angular momentum operator. In the next section, we demonstrate a method for quantizing fields in terms of modes with definite values of the topological charges.

\section{Quantization of vector optical vortices}

To consider the quantization procedure for optical fields with a spatially inhomogeneous polarization distribution, we consider the expansion of the vector potential in Laguerre-Gaussian modes. This representation, already introduced in the classical description given by Eq. \eqref{MultiA}, extends naturally to the quantum case:

\BY
	\hat{\bold{A}}(\rho,\phi) = \sum_{s=-\infty}^{+\infty} \hat{a}_R^s \bold{e}_R   U_s(\rho) e^{is\phi}
	+ \sum_{l=-\infty}^{+\infty} \hat{a}_L^l \bold{e}_L   U_l(\rho) e^{il\phi}\label{MMode},\;\;\;
\EY
where the annihilation operators in modes with right and left circular polarizations and a definite value of the topological charge, $\hat{a}_R^s$ and $\hat{a}_L^l$, replace the corresponding classical amplitudes. The indices $R$ and $L$ correspond to right and left circular polarizations. This expression is a superposition of scalar vortex modes in the two orthogonal polarizations, which makes it possible to represent an arbitrary field in this form. The creation and annihilation operators used here obey the bosonic commutation relations
\BY
\[\hat{a}_\lambda^s(t),\hat{a}_{\lambda^\prime}^{l\, \dag}(t^\prime)\]=\delta_{\lambda,\lambda^\prime}\delta_{s,l}\delta(t-t^\prime),\label{comm}
\EY
where $l$ and $s$ are topological charges, and the indices $\lambda'$ and $\lambda \in \{R, L\}$ denote the polarization.

The expansion in Eq. \eqref{MMode} is essentially a representation of the vector potential as a superposition of modes with an axially symmetric phase distribution. To obtain an expansion in terms of modes with an axially symmetric polarization distribution, we reverse the sign of the summation index in the first term. For each mode with a given index $l$ in the left circular polarization, we then choose the corresponding term with index $-l$ in the right circular polarization. The expression for the vector potential then takes the form
\BY
	\hat{\bold{A}}(\rho,\phi) = \sum_{l=-\infty}^{+ \infty} \left(\hat{a}_R^{-l} U_{-l} (\rho) e^{- i l \phi} \bold{e}_R +\hat{a}_L^l U_l(\rho) e^{i l \phi} \bold{e}_L \right).\;\;\;\;
\EY
We use the property $U_l(\rho)=U_{-l}(\rho)$ and go over to the polarization representation in the basis given by Eq. (\ref{basis}). The expression then takes the form:
\BY
	\hat{\bold{A}}(\rho,\phi) = \sum_{w = - \infty}^{\infty} \left[ \hat{a}_{RP}^w \bold{e}_{RP}^{w}(\phi) + \hat{a}_{AP}^w \bold{e}_{AP}^{w}(\phi) \right] U_w(\rho),\;\;
\EY
where the angular dependence of the polarization vectors $\mathbf{e}_{RP}^{w}$ and $\mathbf{e}_{AP}^{w}$ makes this expansion represent a superposition of azimuthally symmetric vector vortices. Thus, for a field quantized in terms of polarization and the topological charge $l$, one can pass to polarization-topological quantization in terms of the helical topological charge $w$. In this quantization terms, the Hilbert space is not the tensor product of the two-dimensional polarization space and the multidimensional space of states with a definite topological charge $l$. Instead, it is a single space in which the polarization and phase structures of the states are intertwined.
The corresponding photon annihilation operators in the radial and azimuthal modes of vector vortex with a helical topological charge $w$ are determined by original annihilation operators for modes of circular polarization with a definite charge $l$:
\BY
	\hat{a}_{RP}^w &= \frac{1}{\sqrt{2}} \eval{\left( \hat{a}_R^{-l} +  \hat{a}_L^l \right)}_{l=w},
	\nonumber
	\\
	\hat{a}_{AP}^w &= \frac{i }{\sqrt{2}} \eval{\left( \hat{a}_R^{-l} - \hat{a}_L^l \right)}_{l=w}.
\EY

Thus, a optical vector vortex with azimuthally symmetric polarization represents a superposition of two scalar vortex components with orthogonal polarizations and opposite topological charge values.

Using the commutation relations for the annihilation operators in modes with a definite topological charge, Eq. \eqref{comm}, we obtain analogous commutation relations for the annihilation operators in modes with a definite $w$:
\begin{eqnarray}
	[\hat{a}_{\lambda'}^{w'}(t),\hat{a}_{\lambda}^{w\, \dagger}(t^\prime)]=\delta_{\lambda,\lambda^\prime} \delta_{w',w}\delta(t-t^\prime),
\end{eqnarray}
where $w$ and $w'$ are the helical topological charges, and the indices $\lambda'$ and $\lambda \in \{RP, AP\}$ denote the polarization states. The commutation relations given above for operators in modes with a definite helical topological charge $w$, as well as for operators in modes with a definite topological charge $l$, are valid only for longitudinal scales smaller than the Rayleigh range of the beam, $z \ll z_R$, that is, near the beam waist as was shown in Ref. \cite{Vashukevich2022high}.

Thus, we have shown that a field quantized in terms of two spatially homogeneous polarizations and the topological charge $l$ can be quantized in terms of helical topological charge $w$. For each value of the helical topological charge, there exist two orthogonal polarization states. In this case, the polarization states are also spatially inhomogeneous and depend on $w$. That is, the quantization is performed in the basis of polarization-phase states.

\section{Conclusion}

This paper develops a gauge-invariant formalism for describing the spin and orbital angular momenta of optical vortices. This formalism allows one to consistently account for the nonparaxial correction and for the nonlocal nature of the transverse part of the vector potential. It is shown that a rigorous separation of the total angular momentum of the electromagnetic field into spin and orbital components is possible only if gauge invariance is properly taken into account. The additional correction term that arises in this case eliminates the nonphysical contributions associated with the longitudinal component of the potential. It also restores the equality between the total angular momentum and the sum of its spin and orbital parts in terms of integrated quantities.

For scalar optical vortices described by Laguerre-Gaussian modes with a homogeneous polarization structure, analytical expressions are obtained for the projections of the spin and orbital angular momenta as asymptotic series in the nonparaxiality parameter. It is shown that direct summation of these series leads to sign-definite contributions that have no clear physical interpretation. These contributions, however, are eliminated when the correction  introduced above is taken into account. The use of Borel regularization makes it possible to assign physical meaning to the corresponding expansions and to obtain the correct paraxial limit. In this limit, the orbital angular momentum is determined by the topological charge, and the spin angular momentum by the helicity of the field. Thus, the well-known result for scalar vortices is reproduced without discarding any of the nonparaxial corrections, but only by properly accounting for gauge invariance and regularization of the divergent series.

For vector optical vortices with a spatially inhomogeneous polarization structure, the formalism developed above is generalized to the basis of radial and azimuthal polarizations. We have defined the helical topological charge, which characterizes the number of complete rotations of the polarization vector when encircling the optical axis. We have also shown that the orbital angular momentum of such beams splits into two qualitatively different components. The first is polarization-independent and arises from the common phase singularity. The second is polarization-dependent and is proportional to the product of the helicity and the helical topological charge. This separation reflects two different origins of the orbital angular momentum of the field. The phase singularity in the transverse profile of the fields gives rise to the spin-independent part of the angular momentum, whereas the rotation of the polarization basis vectors and the polarization singularity generate the spin-dependent part.

We have also generalized the calculations to an arbitrary multimode vector field. We have shown that the helicity $\sigma$ and topological charges $l$ and $w$ are naturally expressed in terms of the average topological charges and the superposition coefficients of the orthogonal polarization components. The orbital angular momentum also contains two contributions determined by different topological charges.

The quantization procedure for optical fields with a spatially inhomogeneous polarization distribution has been considered. It is shown that the expansion of the vector potential in terms of modes with circular polarizations and definite topological charges can be rewritten as an expansion in terms of modes of radial and azimuthal polarization with a helical topological charge. Importantly, the Hilbert space of such states does not, in general, factorize into polarization and topological parts. This provides additional possibilities for encoding and processing quantum information, as well as for constructing protocols of quantum communication and quantum memory for multidimensional states.

This work was supported by the RSF (grant 26-72-10046). 
\appendix
    \section{Orbital angular momentum of vector vortex fields}
    We assume that the vector potential of a single-mode vector vortex field is given by Eq. (\ref{eq: Pot2}) in the basis of radial and azimuthal polarizations, with coefficients $\alpha_{RP}^w$ and $\alpha_{AP}^w$. To calculate gauge-invariant quantities correctly, it is necessary to compute the n-fold application of the transverse Laplacian to the divergence of the vector potential:
	
	\BY
		&&	\left(\frac{\{\partial_j \partial_j\}^\perp}{k_0^2}\right)^{\!n}
			\partial_i A^{vec}_i(\rho,\phi) = 
		\nn	\\
		&&	=\frac{(-4)^n}{\rho}
			\frac{1}{k_0^{2+2n} \rho_0^{2+2n}}\,(1+n)!\,
			\rho_0^2 \, U_w(\rho)\ L_{n+1}^{(w-1)}\!\left(\frac{\rho^2}{\rho_0^2}\right)\nn
			\\
		&&	\times\Big[
			\cos((1-w)\phi)\alpha_{\mathrm{RP}}^w-2i(\sin((1-w)\phi))\alpha_{\mathrm{AP}}^w		\Big].
	\EY
	
    For vector vortices, the analytical expression for the gauge-invariant potential $\mathcal{A}^{vec}$ makes it possible to integrate directly the $z$-components of the angular momentum densities over all space without truncating the series in Eq. (20). This is similar to the scalar-vortex case. As a result, the projections $S_z^{inv}$ and $L_z^{inv}$ can be represented as asymptotic expansions in the small parameter $\xi=\frac{\sqrt{2}}{k_0^2\rho_0^2}$:
\BY
	&&	S_z^{inv} \propto \sigma\left(1 + \sum_{n=0}^{\infty}
\frac{(-1)^{n+1}\, (1 + w + n)! \, \xi^{n+1}}{w!}\right),\\
			&&		L_z^{inv} \propto w\sigma\(1  +  \sum_{n=0}^{\infty}
			\frac{(-1)^{n+1}\, (1+ w + n)! \xi^{n+1} \, }{w!}\)
			\\
			&&+ \sigma\,(w+1)\sum_{n=0}^{\infty}
			\frac{ (1 + w+2n)! (1+(2+w+2n)\xi)\xi^{2n+1}}{2w!},\;\;\nn
\EY
	where $\sigma=|\alpha_R|^2-|\alpha_L|^2=|\frac{\alpha_{RP}^w+i\alpha_{AP}^w}{\sqrt{2}}|^2-|\frac{\alpha_{RP}^w-i\alpha_{AP}^w}{\sqrt{2}}|^2$.

    The nonphysical contribution from the divergent sign-definite series is eliminated by taking the correction term into account. We write expressions for the total angular momentum and for the correction term, following the procedure of Eqs. \eqref{Glob}--\eqref{Belin}:
\BY
			&&J_z^{inv} \propto \frac{w\sigma+\sigma}{2}\(1+ \sum_{n=0}^{\infty}\frac{ (-1)^n (w+n)!\,\xi^n}{w!}\)
			\\
			&&+ \frac{\sigma}{2} \sum_{n=0}^{\infty}
			\frac{(-1)^{n+1}(w+n)!\big((1+n)(1+w+n)\xi + n\big)\xi^n}{w!},\nn\\
&&			G_z^{inv} \propto\frac{w\sigma+\sigma}{2}\Bigg(1+\nn\\
&&  \sum_{n=0}^{\infty}	\frac{(-1)^{n+1}(w+n)!\big(1 + 2(1+n+w)\xi\big)\xi^n}{w!}\Bigg)
		\nn	\\
			&&+ \frac{\sigma}{2} \sum_{n=0}^{\infty}
			\frac{(-1)^n (w+n)!\big((1+n)(1+w+n)\xi + n\big)\xi^n}{w!}
			\\
			&&+ \frac{w\sigma+\sigma}{2} \sum_{n=0}^{\infty}
			\frac{(1+w+2n)!\big(1 + (2+2n+w)\xi\big)\xi^{2n+1}}{w!}.\nn\;\;\;\;
\EY
	
    We note that the sign-definite series in $G_z^{inv}$ coincides with that in $L_z^{inv}$. Thus, as for scalar vortices, taking the correction into account makes it possible to eliminate the nonphysical contributions to the orbital angular momentum.
	
    After rearranging the contributions, the projections of the spin and orbital angular momenta onto the propagation axis take the following form:
	
\BY
	&&	\tilde{L}_z^{inv} = \frac{w\sigma}{2} \left( 1 + \sum_{n=0}^{\infty}
		\frac{(-1)^n (w+n)!\, \xi^n}{w!} \right),\\
	&&	\tilde{S}_z^{inv} = \frac{\sigma}{2} \left( 1 + \sum_{n=0}^{\infty}
		\frac{(-1)^n (w+n)!\, \xi^n}{w!}\right.\\
	&&\left.	+ \sum_{n=0}^{\infty} \frac{(-1)^n (w+n)!\,
			\big((1+n)(1+w+n)\xi + n\big)\, \xi^n}{w!} \right).\nn
\EY
	
    Applying Borel regularization to the series in the expressions for $\tilde{L}_z^{inv}$ and $\tilde{S}_z^{inv}$, we obtain:
	\BY
	&&	\sum_{n=0}^{\infty} \frac{(-1)^n (w+n)!\, \xi^n}{w!}
	\equiv
		\frac{e^{1/\xi}}{\xi^{w+1}} \int_0^{\infty} \frac{e^{-t/\xi}}{t^{w+1}}\, dt,\\
	&&	\sum_{n=0}^{\infty} \frac{(-1)^n (w+n)!\,
			\big((1+n)(1+w+n)\xi + n\big)\, \xi^n}{w!}
	\equiv
		0.\;\;\;\;\;\;\;\;
\EY
    In the limit of an infinitely wide beam waist ($\xi \to 0$), the integral tends to unity:
	\begin{equation}
		\lim_{\xi \to 0} \frac{e^{1/\xi}}{\xi^{w+1}}
		\int_0^{\infty} \frac{e^{-t/\xi}}{t^{w+1}}\, dt = 1.
	\end{equation}
	Then the Eqs. \eqref{Lvort}-\eqref{Svort} hold for the gauge-invariant projections of the angular momenta.

\end{document}